\pdfoutput=1
\documentclass[12pt,a4paper,titlepage]{article}
\usepackage[utf8]{inputenc}
\usepackage[top=50pt,bottom=50pt,left=68pt,right=66pt]{geometry}
\usepackage{amsmath}
\usepackage{booktabs}
\usepackage{amssymb}
\usepackage[title]{appendix}
\usepackage{mathrsfs}
\usepackage{float}
\usepackage{multirow}
\usepackage{graphicx,caption,subcaption}
\usepackage[space]{grffile}

\begin{document}
\renewcommand{\arraystretch}{1.3}

\makeatletter
\def\@hangfrom#1{\setbox\@tempboxa\hbox{{#1}}%
      \hangindent 0pt%\wd\@tempboxa
      \noindent\box\@tempboxa}
\makeatother

% Underline for text or math

\def\un#1{\relax\ifmmode\@@underline#1\else
        $\@@underline{\hbox{#1}}$\relax\fi}

% Accents and foreign (in text):

\let\under=\unt                 % bar-under (but see \un above)
\let\ced=\ce                    % cedilla
\let\du=\du                     % dot-under
\let\um=\Hu                     % Hungarian umlaut
\let\sll=\lp                    % slashed (suppressed) l (Polish)
\let\Sll=\Lp                    % " L
\let\slo=\os                    % slashed o (Scandinavian)
\let\Slo=\Os                    % " O
\let\tie=\ta                    % tie-after (semicircle connecting two letters)
\let\br=\ub                     % breve
                % Also: \`        grave
                %       \'        acute
                %       \v        hacek (check)
                %       \^        circumflex (hat)
                %       \~        tilde (squiggle)
                %       \=        macron (bar-over)
                %       \.        dot (over)
                %       \"        umlaut (dieresis)
                %       \aa \AA   A-with-circle (Scandinavian)
                %       \ae \AE   ligature (Latin & Scandinavian)
                %       \oe \OE   " (French)
                %       \ss       es-zet (German sharp s)
                %       \$  \#  \&  \%  \pounds  {\it\&}  \dots

% Abbreviations for Greek letters

\def\a{\alpha}
\def\b{\beta}
\def\c{\chi}
\def\d{\delta}
\def\e{\epsilon}
\def\f{\phi}
\def\g{\gamma}
\def\h{\eta}
\def\i{\iota}
\def\j{\psi}
\def\k{\kappa}
\def\l{\lambda}
\def\m{\mu}
\def\n{\nu}
\def\o{\omega}
\def\p{\pi}
\def\q{\theta}
\def\r{\rho}
\def\s{\sigma}
\def\t{\tau}
\def\u{\upsilon}
\def\x{\xi}
\def\z{\zeta}
\def\D{\Delta}
\def\F{\Phi}
\def\G{\Gamma}
\def\J{\Psi}
\def\L{\Lambda}
\def\O{\Omega}
\def\P{\Pi}
\def\Q{\Theta}
\def\S{\Sigma}
\def\U{\Upsilon}
\def\X{\Xi}

% Varletters

\def\ve{\varepsilon}
\def\vf{\varphi}
\def\vr{\varrho}
\def\vs{\varsigma}
\def\vq{\vartheta}

% Calligraphic letters

\def\ca{{\cal A}}
\def\cb{{\cal B}}
\def\cc{{\cal C}}
\def\cd{{\cal D}}
\def\ce{{\cal E}}
\def\cf{{\cal F}}
\def\cg{{\cal G}}
\def\ch{{\cal H}}
\def\ci{{\cal I}}
\def\cj{{\cal J}}
\def\ck{{\cal K}}
\def\cl{{\cal L}}
\def\cm{{\cal M}}
\def\cn{{\cal N}}
\def\co{{\cal O}}
\def\cp{{\cal P}}
\def\cq{{\cal Q}}
\def\car{{\cal R}}
\def\cs{{\cal S}}
\def\ct{{\cal T}}
\def\cu{{\cal U}}
\def\cv{{\cal V}}
\def\cw{{\cal W}}
\def\cx{{\cal X}}
\def\cy{{\cal Y}}
\def\cz{{\cal Z}}

% Fonts

\def\Sc#1{{\hbox{\sc #1}}}      % script for single characters in equations
\def\Sf#1{{\hbox{\sf #1}}}      % sans serif for single characters in equations

                        % Also:  \rm      Roman (default for text)
                        %        \bf      boldface
                        %        \it      italic
                        %        \mit     math italic (default for equations)
                        %        \sl      slanted
                        %        \em      emphatic
                        %        \tt      typewriter
                        % and sizes:    \tiny
                        %               \scriptsize
                        %               \footnotesize
                        %               \small
                        %               \normalsize
                        %               \large
                        %               \Large
                        %               \LARGE
                        %               \huge
                        %               \Huge

% Math symbols

\def\slpa{\slash{\pa}}                            % slashed partial derivative
\def\slin{\SLLash{\in}}                                   % slashed in-sign
\def\bo{{\raise-.3ex\hbox{\large$\Box$}}}               % D'Alembertian
\def\cbo{\Sc [}                                         % curly "
\def\pa{\partial}                                       % curly d
\def\de{\nabla}                                         % del
\def\dell{\bigtriangledown}                             % hi ho the dairy-o
\def\su{\sum}                                           % summation
\def\pr{\prod}                                          % product
\def\iff{\leftrightarrow}                               % <-->
\def\conj{{\hbox{\large *}}}                            % complex conjugate
\def\ltap{\raisebox{-.4ex}{\rlap{$\sim$}} \raisebox{.4ex}{$<$}}   % < or ~
\def\gtap{\raisebox{-.4ex}{\rlap{$\sim$}} \raisebox{.4ex}{$>$}}   % > or ~
\def\TH{{\raise.2ex\hbox{$\displaystyle \bigodot$}\mskip-4.7mu \llap H \;}}
\def\face{{\raise.2ex\hbox{$\displaystyle \bigodot$}\mskip-2.2mu \llap {$\ddot
        \smile$}}}                                      % happy face
\def\dg{\sp\dagger}                                     % hermitian conjugate
\def\ddg{\sp\ddagger}                                   % double dagger
                        % Also:  \int  \oint              integral, contour
                        %        \hbar                    h bar
                        %        \infty                   infinity
                        %        \sqrt                    square root
                        %        \pm  \mp                 plus or minus
                        %        \cdot  \cdots            centered dot(s)
                        %        \oplus  \otimes          group theory
                        %        \equiv                   equivalence
                        %        \sim                     ~
                        %        \approx                  approximately =
                        %        \propto                  funny alpha
                        %        \ne                      not =
                        %        \le \ge                  < or = , > or =
                        %        \{  \}                   braces
                        %        \to  \gets               -> , <-
                        % and spaces:  \,  \:  \;  \quad  \qquad
                        %              \!                 (negative)

\font\tenex=cmex10 scaled 1200

% Math stuff with one argument

\def\sp#1{{}^{#1}}                              % superscript (unaligned)
\def\sb#1{{}_{#1}}                              % sub"
\def\oldsl#1{\rlap/#1}                          % poor slash
\def\slash#1{\rlap{\hbox{$\mskip 1 mu /$}}#1}      % good slash for lower case
\def\Slash#1{\rlap{\hbox{$\mskip 3 mu /$}}#1}      % " upper
\def\SLash#1{\rlap{\hbox{$\mskip 4.5 mu /$}}#1}    % " fat stuff (e.g., M)
\def\SLLash#1{\rlap{\hbox{$\mskip 6 mu /$}}#1}      % slash for no-in sign
\def\PMMM#1{\rlap{\hbox{$\mskip 2 mu | $}}#1}   %
\def\PMM#1{\rlap{\hbox{$\mskip 4 mu ~ \mid $}}#1}       %
\def\Tilde#1{\widetilde{#1}}                    % big tilde
\def\Hat#1{\widehat{#1}}                        % big hat
\def\Bar#1{\overline{#1}}                       % big bar
\def\sbar#1{\stackrel{*}{\Bar{#1}}}             % big bar with star
\def\bra#1{\left\langle #1\right|}              % < |
\def\ket#1{\left| #1\right\rangle}              % | >
\def\VEV#1{\left\langle #1\right\rangle}        % < >
\def\abs#1{\left| #1\right|}                    % | |
\def\leftrightarrowfill{$\mathsurround=0pt \mathord\leftarrow \mkern-6mu
        \cleaders\hbox{$\mkern-2mu \mathord- \mkern-2mu$}\hfill
        \mkern-6mu \mathord\rightarrow$}
\def\dvec#1{\vbox{\ialign{##\crcr
        \leftrightarrowfill\crcr\noalign{\kern-1pt\nointerlineskip}
        $\hfil\displaystyle{#1}\hfil$\crcr}}}           % <--> accent
\def\dt#1{{\buildrel {\hbox{\LARGE .}} \over {#1}}}     % dot-over for sp/sb
\def\dtt#1{{\buildrel \bullet \over {#1}}}              % alternate "
\def\der#1{{\pa \over \pa {#1}}}                % partial derivative
\def\fder#1{{\d \over \d {#1}}}                 % functional derivative
                % Also math accents:    \bar
                %                       \check
                %                       \hat
                %                       \tilde
                %                       \acute
                %                       \grave
                %                       \breve
                %                       \dot    (over)
                %                       \ddot   (umlaut)
                %                       \vec    (vector)

% Math stuff with more than one argument

\def\frac#1#2{{\textstyle{#1\over\vphantom2\smash{\raise.20ex
        \hbox{$\scriptstyle{#2}$}}}}}                   % fraction
\def\half{\frac12}                                        % 1/2
\def\sfrac#1#2{{\vphantom1\smash{\lower.5ex\hbox{\small$#1$}}\over
        \vphantom1\smash{\raise.4ex\hbox{\small$#2$}}}} % alternate fraction
\def\bfrac#1#2{{\vphantom1\smash{\lower.5ex\hbox{$#1$}}\over
        \vphantom1\smash{\raise.3ex\hbox{$#2$}}}}       % "
\def\afrac#1#2{{\vphantom1\smash{\lower.5ex\hbox{$#1$}}\over#2}}    % "
\def\partder#1#2{{\partial #1\over\partial #2}}   % partial derivative of
\def\parvar#1#2{{\d #1\over \d #2}}               % variation of
\def\secder#1#2#3{{\partial^2 #1\over\partial #2 \partial #3}}  % second "
\def\on#1#2{\mathop{\null#2}\limits^{#1}}               % arbitrary accent
\def\bvec#1{\on\leftarrow{#1}}                  % backward vector accent
\def\oover#1{\on\circ{#1}}                              % circle accent

\def\[{\lfloor{\hskip 0.35pt}\!\!\!\lceil}
\def\]{\rfloor{\hskip 0.35pt}\!\!\!\rceil}
\def\Lag{{\cal L}}
\def\du#1#2{_{#1}{}^{#2}}
\def\ud#1#2{^{#1}{}_{#2}}
\def\dud#1#2#3{_{#1}{}^{#2}{}_{#3}}
\def\udu#1#2#3{^{#1}{}_{#2}{}^{#3}}
\def\calD{{\cal D}}
\def\calM{{\cal M}}

\def\szet{{${\scriptstyle \b}$}}
\def\ulA{{\un A}}
\def\ulM{{\underline M}}
\def\cdm{{\Sc D}_{--}}
\def\cdp{{\Sc D}_{++}}
\def\vTheta{\check\Theta}
\def\fracm#1#2{\hbox{\large{${\frac{{#1}}{{#2}}}$}}}
\def\ha{{\fracmm12}}
\def\tr{{\rm tr}}
\def\Tr{{\rm Tr}}
\def\itrema{$\ddot{\scriptstyle 1}$}
\def\ula{{\underline a}} \def\ulb{{\underline b}} \def\ulc{{\underline c}}
\def\uld{{\underline d}} \def\ule{{\underline e}} \def\ulf{{\underline f}}
\def\ulg{{\underline g}}
\def\items#1{\\ \item{[#1]}}
\def\ul{\underline}
\def\un{\underline}
\def\fracmm#1#2{{{#1}\over{#2}}}
\def\footnotew#1{\footnote{\hsize=6.5in {#1}}}
\def\low#1{{\raise -3pt\hbox{${\hskip 0.75pt}\!_{#1}$}}}

\def\Dot#1{\buildrel{_{_{\hskip 0.01in}\bullet}}\over{#1}}
\def\dt#1{\Dot{#1}}

\def\DDot#1{\buildrel{_{_{\hskip 0.01in}\bullet\bullet}}\over{#1}}
\def\ddt#1{\DDot{#1}}

\def\DDDot#1{\buildrel{_{_{\hskip 0.01in}\bullet\bullet\bullet}}\over{#1}}
\def\dddt#1{\DDDot{#1}}

\def\DDDDot#1{\buildrel{_{_{\hskip 
0.01in}\bullet\bullet\bullet\bullet}}\over{#1}}
\def\ddddt#1{\DDDDot{#1}}

\def\Tilde#1{{\widetilde{#1}}\hskip 0.015in}
\def\Hat#1{\widehat{#1}}

% Aligned equations

\newskip\humongous \humongous=0pt plus 1000pt minus 1000pt
\def\caja{\mathsurround=0pt}
\def\eqalign#1{\,\vcenter{\openup2\jot \caja
        \ialign{\strut \hfil$\displaystyle{##}$&$
        \displaystyle{{}##}$\hfil\crcr#1\crcr}}\,}
\newif\ifdtup
\def\panorama{\global\dtuptrue \openup2\jot \caja
        \everycr{\noalign{\ifdtup \global\dtupfalse
        \vskip-\lineskiplimit \vskip\normallineskiplimit
        \else \penalty\interdisplaylinepenalty \fi}}}
\def\li#1{\panorama \tabskip=\humongous                         % eqalignno
        \halign to\displaywidth{\hfil$\displaystyle{##}$
        \tabskip=0pt&$\displaystyle{{}##}$\hfil
        \tabskip=\humongous&\llap{$##$}\tabskip=0pt
        \crcr#1\crcr}}
\def\eqalignnotwo#1{\panorama \tabskip=\humongous
        \halign to\displaywidth{\hfil$\displaystyle{##}$
        \tabskip=0pt&$\displaystyle{{}##}$
        \tabskip=0pt&$\displaystyle{{}##}$\hfil
        \tabskip=\humongous&\llap{$##$}\tabskip=0pt
        \crcr#1\crcr}}

% Some defs

\def\eV{\,{\rm eV}}
\def\keV{\,{\rm keV}}
\def\MeV{\,{\rm MeV}}
\def\GeV{\,{\rm GeV}}
\def\TeV{\,{\rm TeV}}
\def\sv{\left<\sigma v\right>}
\def\({\left(}
\def\){\right)}
\def\cm{{\,\rm cm}}
\def\K{{\,\rm K}}
\def\kpc{{\,\rm kpc}}
\def\beq{\begin{equation}}
\def\eeq{\end{equation}}
\def\bea{\begin{eqnarray}}
\def\eea{\end{eqnarray}}

% New commands

\newcommand{\be}{\begin{equation}}
\newcommand{\ee}{\end{equation}}
\newcommand{\nbe}{\begin{equation*}}
\newcommand{\nee}{\end{equation*}}

\newcommand{\fr}{\frac}
\newcommand{\lb}{\label}

\thispagestyle{empty}

{\hbox to\hsize{
\vbox{\noindent June 2023 \hfill IPMU23-0011 \\
\noindent revised version \hfill }}

\noindent
\vskip2.0cm
\begin{center}

{\large\bf Production of primordial black holes \\ in improved E-models of inflation}

\vglue.3in

Daniel Frolovsky~${}^{a}$ and Sergei V. Ketov~${}^{a,b,c,\#}$ 
\vglue.3in

${}^a$~Interdisciplinary Research Laboratory, 
Tomsk State University, Tomsk 634050, Russia\\
${}^b$~Department of Physics, Tokyo Metropolitan University, Hachioji, Tokyo 192-0397, Japan \\
${}^c$~Kavli Institute for the Physics and Mathematics of the Universe (WPI),
\\The University of Tokyo Institutes for Advanced Study,  Chiba 277-8583, Japan\\
\vglue.2in

${}^{\#}$~ketov@tmu.ac.jp
\end{center}

\vglue.3in

\begin{center}
{\Large\bf Abstract}  
\end{center}
\vglue.2in

\noindent The E-type $\alpha$-attractor models of single-field inflation were generalized further in order to accommodate production of
primordial black holes (PBH) via adding a near-inflection point to the inflaton scalar potential at smaller scales, in good agreement with measurements of the cosmic microwave background (CMB) radiation. A minimal number of new parameters was used but their fine-tuning was maximized in order to  increase possible masses of PBH formed during an ultra-slow-roll phase leading to a large enhancement of the power spectrum of scalar (curvature) perturbations by 6 or 7 orders of magnitude against the power spectrum of perturbations observed in CMB. It was found that extreme fine-tuning of the parameters in our models can lead to a formation of Moon-size PBH with the masses up to $10^{26}$ g, still in agreement with CMB observations. Quantum corrections are known to lead to the perturbative upper bound on the amplitude of large scalar perturbations responsible for PBH production. The quantum (one-loop) corrections in our models were found to be suppressed by one order of magnitude for PBH with the masses of approximately $10^{19}$ g, which may form the whole dark matter in the Universe.

\newpage

\section{Introduction}

The paradigm of cosmological inflation proposed by Guth and Linde \cite{Guth:1980zm,Linde:1981mu} is a possible answer to internal problems of the standard (Einstein-Friedmann) cosmology. Inflation as the amplifier of curvature perturbations is also a possible origin of seeds of structure formation in the early Universe \cite{Mukhanov:2005sc,Liddle:2000cg}. The existence of inflation is supported by measurements of the cosmic microwave background (CMB) radiation \cite{Planck:2018jri,BICEP:2021xfz,Tristram:2021tvh}.

The CMB measurements  give a small window into high energy physics in the form of the power spectrum of perturbations on limited (CMB) scales and lead to important  restrictions on viable inflationary models but do not allow a reconstruction of the unique underlying inflationary model. There are uncertainties even among the simplest single-field (quintessence) models of inflation in the form of the (canonical) inflaton
scalar potential. Those uncertainties can be exploited to win more room for the single-field inflationary models by extending flexibility in their predictions about inflation, while simultaneously describing formation of primordial black holes (PBH) from large scalar perturbations  at lower (than CMB) scales \cite{Novikov:1967tw,Hawking:1971ei}. A standard mechanism of PBH formation is given by adding 
a near-inflection point to the inflaton potential \cite{Ivanov:1994pa,Garcia-Bellido:1996mdl,Garcia-Bellido:2017mdw,Germani:2017bcs}. The observational constraints on PBH are also subject to uncertainties~\cite{Carr:2020gox,Escriva:2022duf} even in the limited context of the single-field models of inflation with a near-inflection point  \cite{Karam:2022nym}. Therefore, it makes sense to investigate the amount of flexibility against fine-tuning in the well-motivated models of inflation and PBH production on the case-by-case basis. 

In this paper we revisit the $\a${\it -attractor} (single-field) inflationary models introduced in Refs.~\cite{Kallosh:2013hoa,Galante:2014ifa} and propose new generalizations of them. The simplest attractor model with $\a=1$ is given by the famous Starobinsky model \cite{Starobinsky:1980te} motivated by gravitational interactions, with inflaton as the Nambu-Goldstone boson associated with spontaneous breaking of scale invariance \cite{Ketov:2012jt}.  After taking into account the CMB constraints alone, the Starobinsky model has no free parameters and leads to a sharp prediction for the tensor power spectrum tilt of the gravitational waves induced by inflation. The (canonical) inflaton scalar potential in the Starobinsky model reads
\begin{equation} \label{starp}
V(\phi) = \fracmm{3}{4} M^2_{\rm Pl}M^2\left[ 1- \exp\left(-\sqrt{\frac{2}{3}}\phi/M_{\rm Pl}\right)\right]^2 \equiv V_0 ( 1 - 2y) +
 {\cal O}(y^2),~ y = \exp\left(-\sqrt{\frac{2}{3}}\phi/M_{\rm Pl}\right),
\end{equation}
where $M\sim 10^{-5}M_{\rm Pl}$ is the inflaton mass and $M_{\rm Pl}$ is the reduced Planck mass.  This leads to the Mukhanov-Chibisov formula \cite{Mukhanov:1981xt} 
for the tilt of scalar perturbations, $n_s\approx 1- 2/N_e$, and the tensor-to-scalar ratio  $r\approx 12/N_e^2$, where $N_e$ 
 is the number of e-folds measuring the duration of inflation, in very good agreement with the
CMB measurements \cite{Planck:2018jri,BICEP:2021xfz,Tristram:2021tvh},
\be \lb{ns}
n_s = 0.9649 \pm 0.0042 \quad (68\%~{\rm C.L.})~,\quad r < 0.032\quad (95\%~{\rm C.L.})~~,
\ee
for $N_{e}=55\pm 10$. Actually, only the leading exponent in Eq.~({\ref{starp}), i.e. the term linear in $y$, is relevant for the tilts during slow-roll inflation predicted by the Starobinsky model. This simple observation allows one to extend the Starobinsky model by a single new parameter $\a>0$ entering the new variable $y$ as \cite{Kallosh:2013hoa,Galante:2014ifa} 
\begin{equation}  \lb{newy}
y = \exp\left(-\sqrt{\fracm{2}{3\a}}\phi/M_{\rm Pl}\right)~,
\end{equation}
without changing the Mukhanov-Chibisov formula for $n_s$. However, the tensor-to-scalar ratio changes as 
\be \lb{ralpha}
r \approx \fracmm{12\a}{N_e^2}~,
\ee
thus adding more flexibility against future measurements of $r$. 

The scalar potentials of the simplest $\a$-attractor models fall into two classes: the simplest {\it E-models} have the scalar potential as in
Eq.~(\ref{starp}) but with the variable $y$ defined by Eq.~(\ref{newy}), whereas the simplest {\it T-models} have the scalar potential 
\be  \lb{gre1}
V_{\rm T}(\phi) = V_0 \tanh^2 \fracmm{\phi/M_{\rm Pl}}{\sqrt{6\a}}\equiv V_0\tilde{r}^2~~,\quad \tilde{r}=\tanh \fracmm{\phi/M_{\rm Pl}}{\sqrt{6\a}}\equiv \tanh \varphi~~,
\ee
where the scale of inflation is given by $V_0$.  Further generalizations are also possible, while keeping the predictions for the cosmological tilts $n_s$ and $r$ during inflation. For example, the generalized T-models with the scalar potential  $V_{\a,f}(\varphi) = f^2 (\tilde{r})$ and a monotonically increasing (during slow roll) function $f(\tilde{r})$ were used in Refs.~\cite{Dalianis:2018frf,Iacconi:2021ltm,Braglia:2022phb} for engineering a near-inflection point in the potential and PBH production, see also Ref.~\cite{Frolovsky:2023xid}. The T-model potentials can also be made periodic by changing their global shape after replacing the function $\tilde{r}(\varphi)$  in Eq.~(\ref{gre1})  by  the periodic (Jacobi) elliptic function  ${\rm sn}(\varphi\left|\right.k)$ with the elliptic modulus  $0<k^2< 1$, in the limit
\be \lb{jacobi} 
{\rm sn}(\varphi \left|\right. k) \approx \tanh \varphi \quad {\rm for} \quad k^2\to 1~~, 
\ee  
thus combining theoretically attractive features of chaotic inflation and natural inflation.

The power spectrum of scalar perturbations in some generalized T-models with a near-inflection point for PBH production and the related spectrum of induced gravitational waves were derived in Ref.~\cite{Frolovsky:2022ewg}. Similar results for the E-models were obtained  in
Ref.~\cite{Frolovsky:2022qpg}. 

The scalar potential used in Ref.~\cite{Frolovsky:2022qpg} has the form
\begin{equation} \label{fkspot}
V(\phi) = \fracmm{3}{4} M^2_{\rm Pl}M^2\left[ 1- y + y^2(\b - \g y)\right]^2~,
\end{equation}
with $y$ defined by Eq.~(\ref{newy}) and the new (positive) parameters $\b$ and $\g$ needed for engineering a near-inflection point.
The ansatz (\ref{fkspot}) was motivated in Ref.~\cite{Frolovsky:2022ewg} by small values of $y$ during slow-roll inflation, keeping  agreement with CMB measurements \cite{Ivanov:2021chn}, and having a near-inflection point at lower scales for certain (fine-tined) values of the new parameters, which was needed for viable inflation and PBH production.~\footnote{Similar models were proposed and studied in Ref.~\cite{Mishra:2019pzq}.}

As was demonstrated in Ref.~\cite{Frolovsky:2022qpg}, the potential (\ref{fkspot}) leads to PBH production with the PBH masses corresponding to the asteroid-size black holes in the small mass window where those PBH may form the whole dark matter in the Universe. However, in Ref.~\cite{Frolovsky:2022qpg}, it was achieved at the price of having the agreement with the CMB value of $n_s$ outside $1\s$ (though within $2 \sigma$), with fine tuning the parameters, and ignoring quantum corrections. Actually, the higher the masses of the PBH produced, the lower (in red) the scalar tilt $n_s$ becomes. Therefore, the following questions arise:
\begin{itemize}
\item Is it possible to reach the perfect (within $1\s$) agreement with the CMB value of $n_s$ in the E-type models of inflation and PBH formation?
\item Is it possible to increase the PBH masses beyond the asteroid-size?
\item Is it possible so suppress (1-loop) quantum corrections?
\end{itemize}

In this communication we improve our earlier findings about E-models in order to achieve the perfect agreement (inside $1\s$) with the observed values of the CMB tilts, reconsider PBH formation and comment on (one-loop) quantum corrections in our new E-type models.

\section{More general E-models}

At least two extra parameters are needed, as in Eq.~(\ref{fkspot}), for engineering a near-inflection point. However, the parameters in
Eq.~(\ref{fkspot}) do not have clear meaning, so that it is useful to replace them by other (related) parameters as 
\cite{Iacconi:2021ltm,Frolovsky:2022qpg}
\begin{equation} \lb{newp2}
\beta=\fracmm{1}{1-{\xi}^2} \exp \left[\sqrt{\fracmm{2}{3 \alpha}} \fracmm{\phi_i}{M_{\mathrm{Pl}}}\right], \quad\gamma=\fracmm{1}{3\left(1-\xi^2\right)} \exp \left[2 \sqrt{\fracmm{2}{3\alpha}} \fracmm{\phi_i}{M_{\mathrm{PI}}}\right].
\end{equation}
The two real parameters $\left(\phi_i, \xi\right)$ have the clear meaning: when $\xi=0$, the potential has the inflection point at $\phi=\phi_i$ only; when $0<\xi \ll 1$, the potential has also a local minimum (dip) and a local maximum (bump) at $y_{\mathrm{ext.}}^{\pm}$, respectively,  while both extrema are equally separated from the inflection point,
\be \lb{sep} y_{\text {ext. }}^{ \pm}=y_i(1 \pm \xi)~,\quad y_i=\exp\left[-\sqrt{\fracmm{2}{3\alpha}} \fracmm{\phi_i}{M_{\mathrm{Pl}}}\right]~.
\ee

Analyticity of the inflaton potential with respect to $y$ (near $y=0$), responsible for the infinite plateau at $y\to +\infty$, can be relaxed by adding {\it negative} powers of $y$. Adding them can be motivated by supergravity cosmology \cite{Aldabergenov:2023yrk} and string cosmology \cite{Cicoli:2023opf}, where such terms naturally appear. For our purposes, we use this resource in the minimalistic manner by adding merely a single term with a negative power to the E-model potential (\ref{fkspot}). We find adding the term proportional to $y^{-2}$ more efficient  than $y^{-1}$ so that we consider only the case of
 \begin{equation} \label{npot}
V(\phi) = \fracmm{3}{4} M^2_{\rm Pl}M^2\left[ 1- y - \theta y^{-2} + y^2(\b - \g y)\right]^2~,
\end{equation}
with the new (third) real parameter $\theta$ that is supposed to be small enough in order to keep agreement with CMB. An impact of the new parameter on the shape of the scalar potential is illustrated on Fig.~\ref{fig1}. The shape needed for inflation and PBH production is not generic, being  found after scanning the parameter space and fine-tuning the parameter values. 

\begin{figure}[h]
\center{\includegraphics[width=0.6\textwidth]{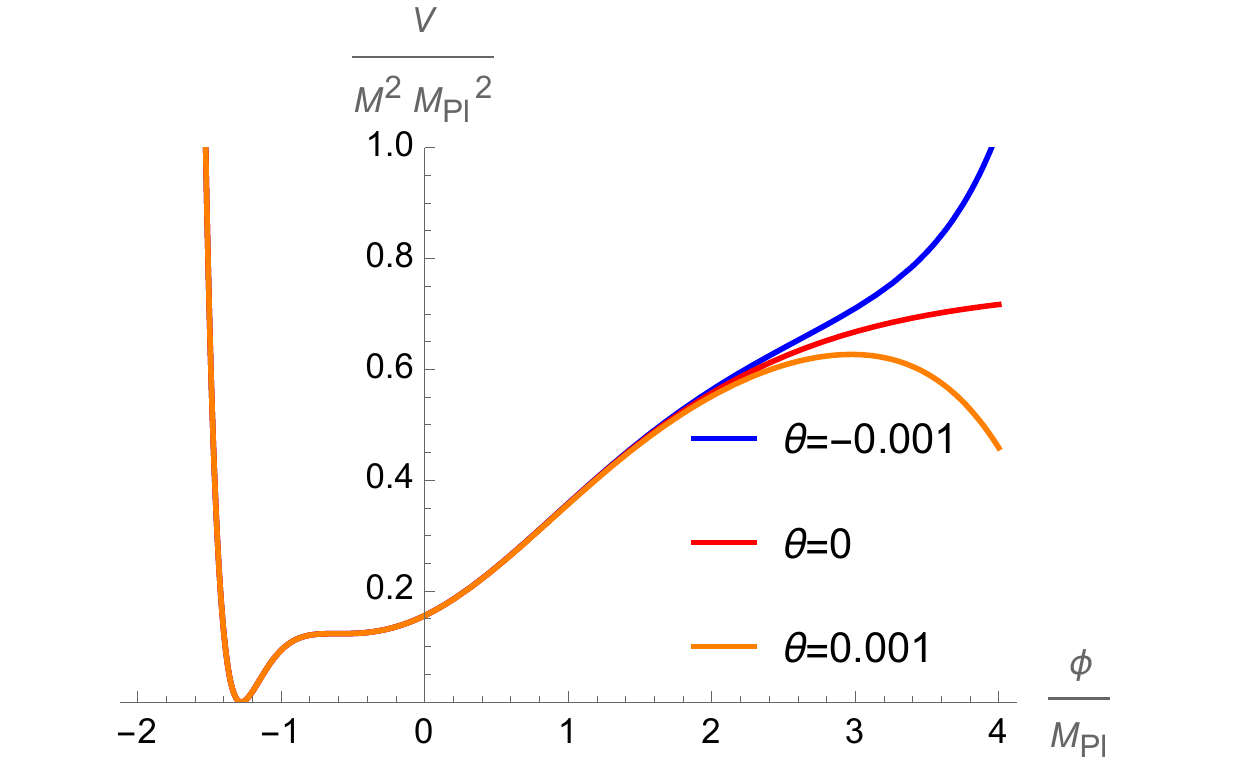}\\}
\caption{The potential (\ref{npot}) with  $\alpha=0.74265$, $\xi =0.01213$  and $\phi_i/M_{\rm Pl}=-0.61115$ for the selected values 
of $\theta$.  }
\label{fig1}
\end{figure}

As is clear from Fig.~1, a negative $\theta$ brings the potential up, and a positive $\theta$ brings the potential down, against the plateau
for $\theta=0$ for large values of $\phi/M_{\rm Pl}$.

The impact of the new parameter $\theta$ on the slow-roll (SR) parameters 
\begin{equation}
	\epsilon_{\rm sr}(\f) =\fracmm{M_{\mathrm{Pl}}^2}{2}\left[\fracmm{V'(\phi)}{V(\phi)}\right]^2~,\qquad 
	\eta_{\rm sr}(\f) =M_{\mathrm{Pl}}^2 \fracmm{V''(\phi)}{V(\phi)}~~,
\end{equation}
is displayed on Fig.~\ref{fig2} that shows the SR phase  in agreement with CMB. The end of this SR phase of inflation is determined by the
condition $\abs{\h}=1$. The standard pivot scale of the horizon exit, when scalar perturbations leave the horizon, is $k_*=0.05~{\rm Mpc}^{-1}$.

\begin{figure}[h]
\begin{minipage}[h]{0.49\linewidth}
\center{\includegraphics[width=1\textwidth]{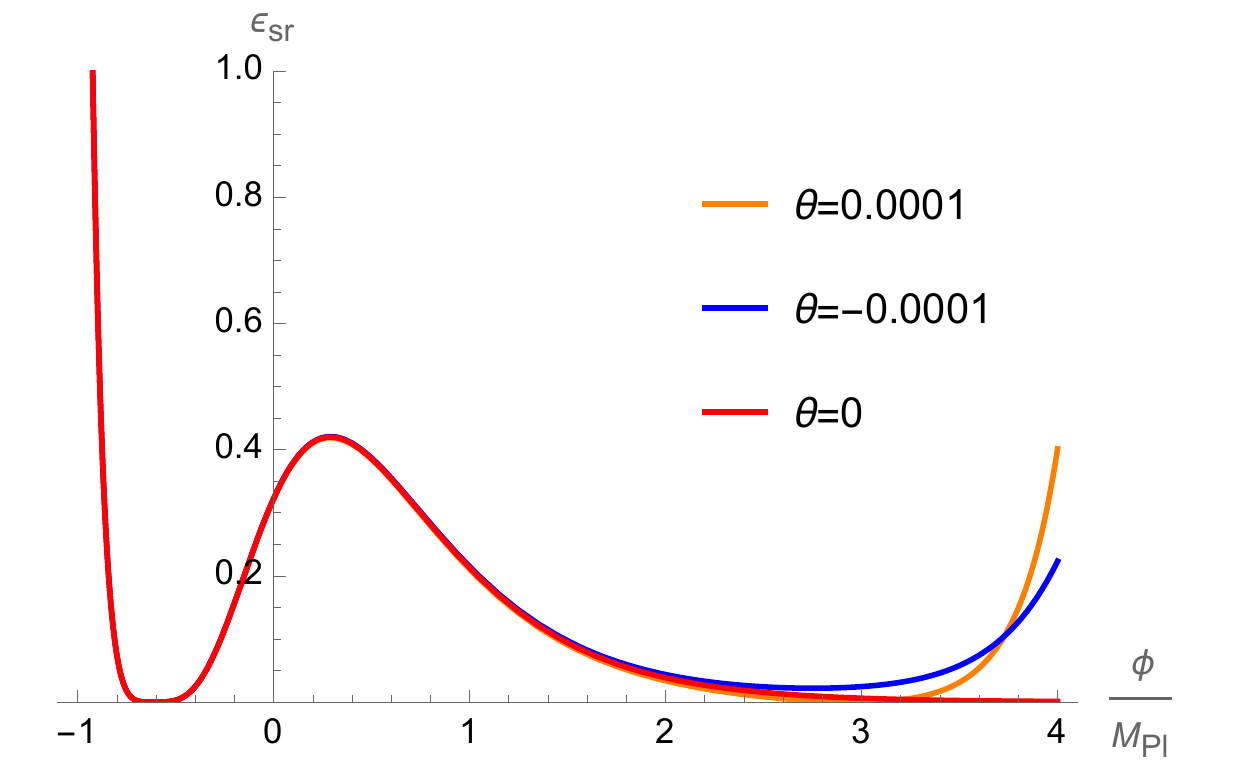}}
\end{minipage}
\hfill
\begin{minipage}[h]{0.49\linewidth}
\center{\includegraphics[width=1\textwidth]{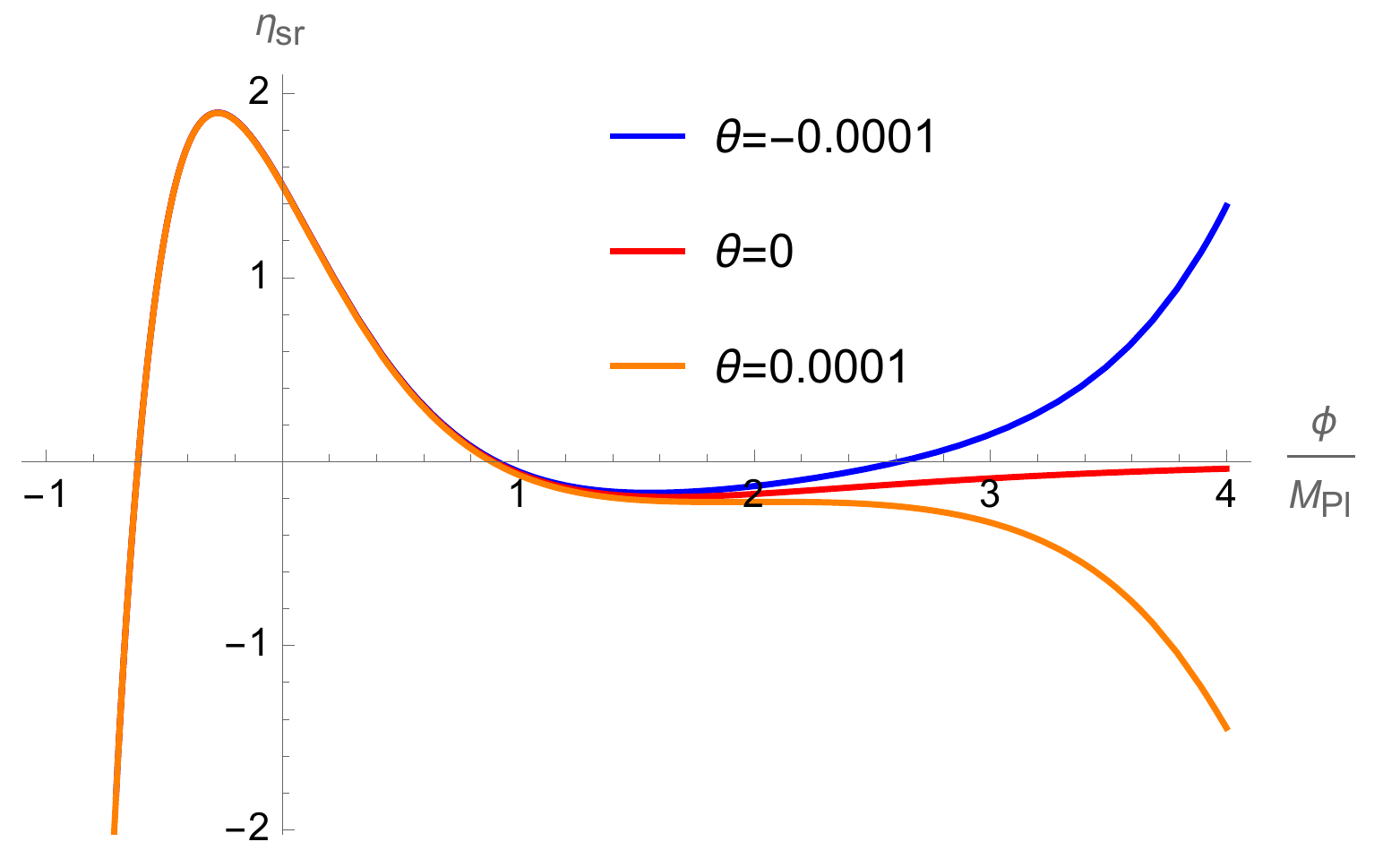}}
\end{minipage}
\caption{The $\phi$-dependence of the slow-roll parameters at $\alpha=0.74265$, $\xi =0.01213$ and $\phi_i/M_{\rm Pl}=-0.61115$ for the selected values of $\theta$.}
\label{fig2}
\end{figure}

The raising tail of the profile of $\e_{\rm sr}(\f)$  near the horizon exit on the left plot of Fig.~\ref{fig2}  increases the tensor-to-scalar ratio $r$,
while the need to increase $n_s$ implies the negative sign of $\theta$, as is clear from the right plot of Fig.~\ref{fig2}. To increase the PBH masses (see the next Section), we can go up to the maximal value $r_{\rm max.}=0.032$ according to Eq.~(\ref{ns}).

Our numerical solutions to the standard equations of motion for the inflaton $\f(t)$ as the function of time $t$ and the Hubble function $H(t)$ with the same values of the parameters $(\a,\x,\f_i)$ for the selected values of $\theta<0$ are shown in Fig.~\ref{fig3}. It shows the duration of inflation,
defined by the length of the plateau, for the different values of $\theta$.

\begin{figure}[h]
\begin{minipage}[h]{0.49\linewidth}
\center{\includegraphics[width=1\textwidth]{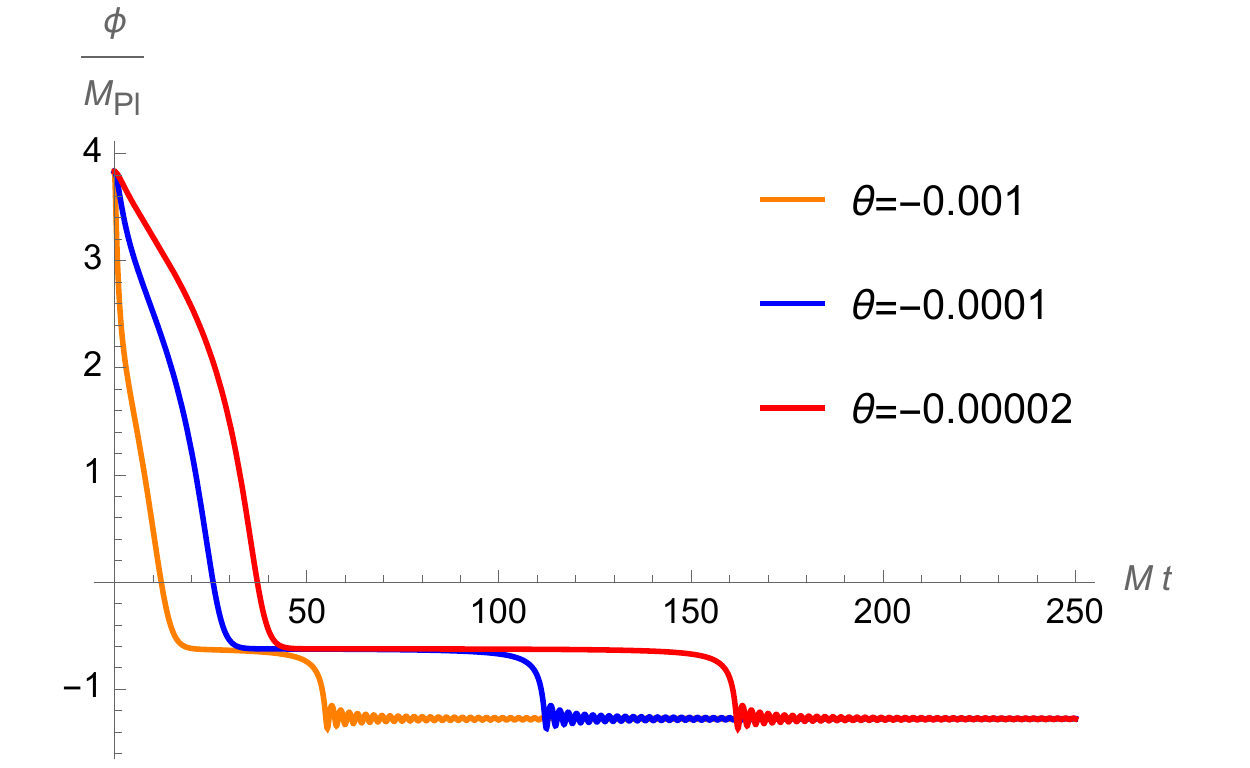}}
\end{minipage}
\hfill
\begin{minipage}[h]{0.49\linewidth}
\center{\includegraphics[width=1\textwidth]{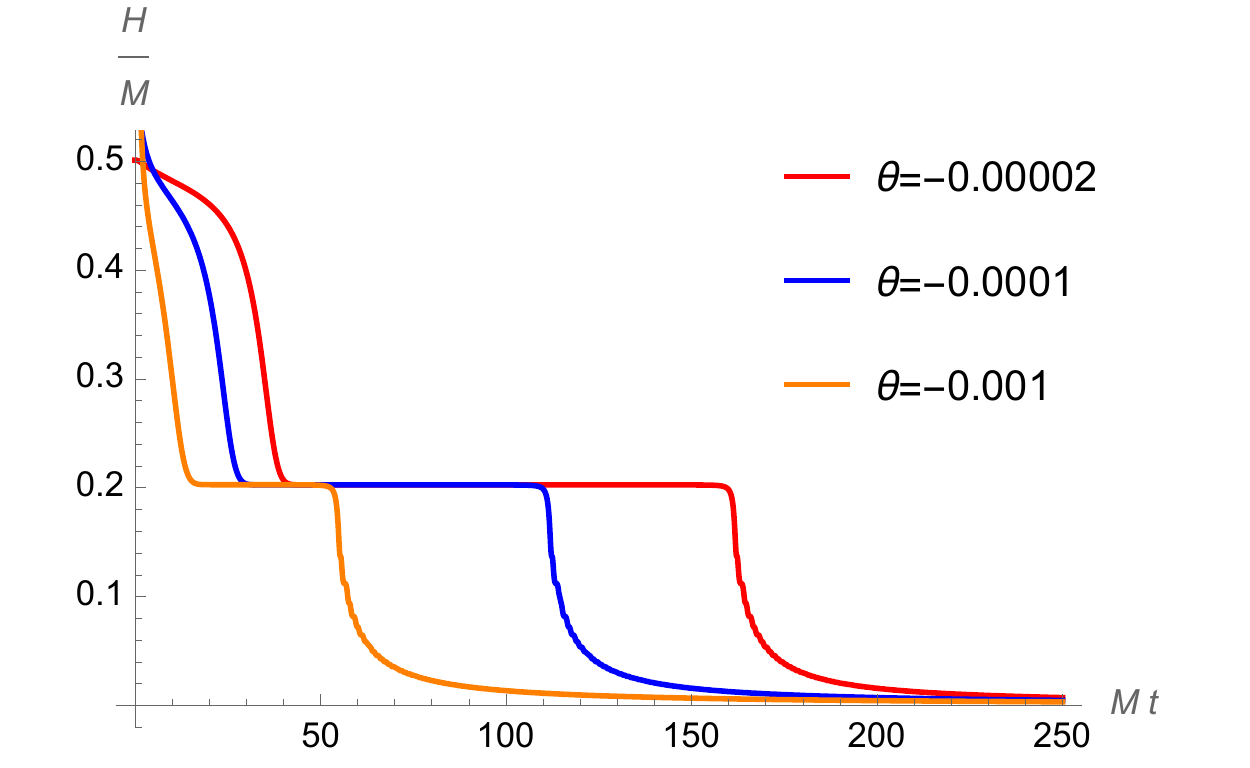}}
\end{minipage}
\caption{The time evolution of the inflaton $\f(t)$ and the Hubble function $H(t)=\dot{a}/a$ (where $a(t)$ is the cosmic factor in the metric
of a spatially flat Friedmann universe) with the  parameters $\alpha=0.74265$, $\xi =0.01213$ and
 $\phi_i/M_{\rm Pl}=-0.61115$, and the horizon exit at $\phi_{\rm in}/M_{\rm Pl}=3.886$. }
\label{fig3}
\end{figure}

The Hubble flow functions are defined by
\begin{equation} \lb{Hflow}
	\epsilon(t)=-\fracmm{\dot H}{H^2}~,\qquad \eta(t)=\fracmm{\dot \epsilon}{H \epsilon}~.
\end{equation}
The (numerically derived) dynamics of them is shown in Fig.~\ref{fig4}.

\begin{figure}[h]
\begin{minipage}[h]{0.49\linewidth}
\center{\includegraphics[width=1\textwidth]{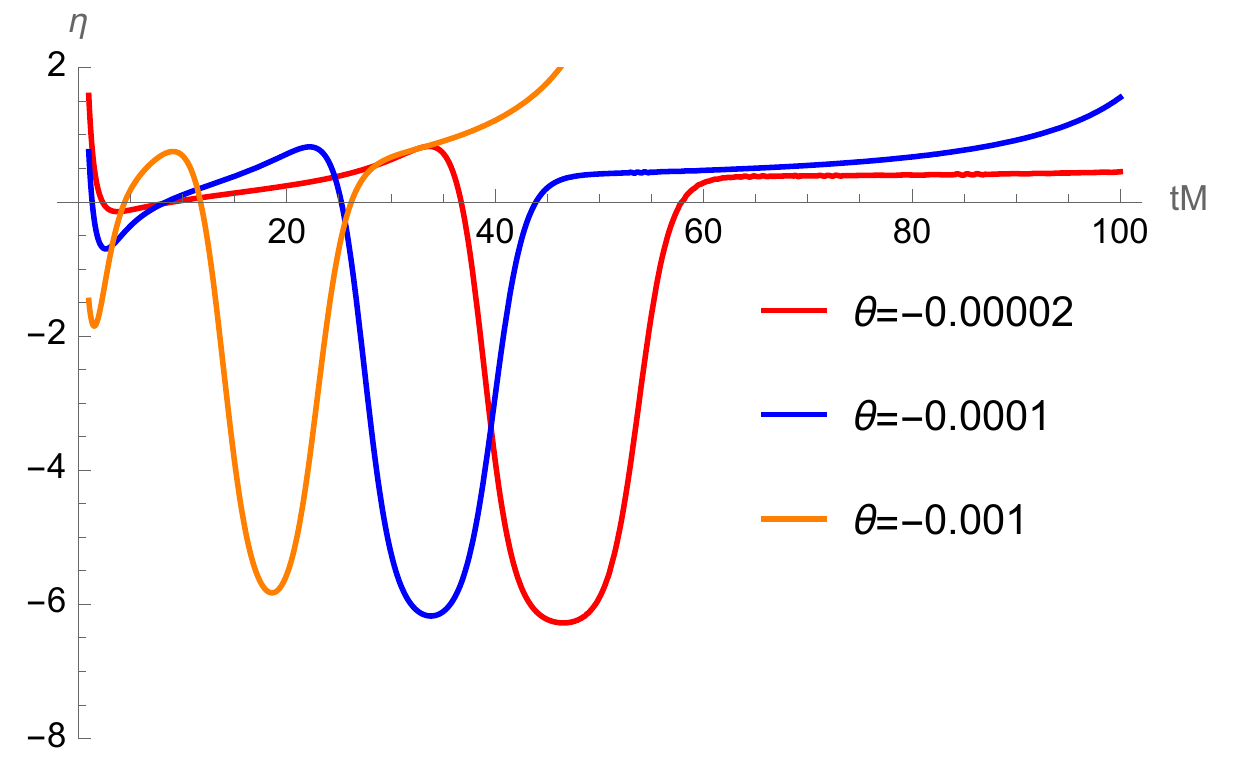}}
\end{minipage}
\hfill
\begin{minipage}[h]{0.49\linewidth}
\center{\includegraphics[width=1\textwidth]{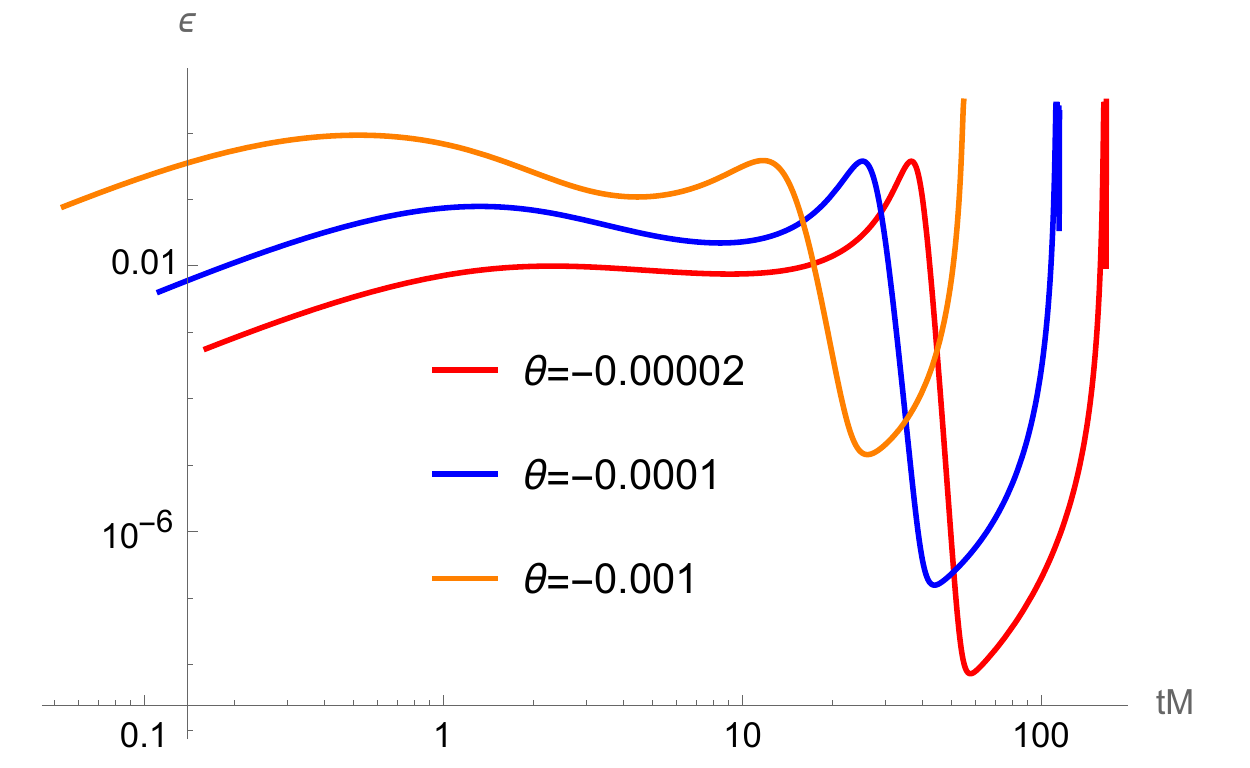}}
\end{minipage}
\caption{The Hubble flow functions $\eta(t)$ and $\epsilon(t)$ with the parameters $\alpha=0.74265$, $\xi =0.01213$ and 
$\phi_i/M_{\rm Pl}=-0.61115$. }
\label{fig4}
\end{figure}

Fig.~\ref{fig4} shows the existence of the ultra-slow-roll (USR) phase \cite{Germani:2017bcs,Dimopoulos:2017ged}, where the first derivative of the inflaton potential and the Hubble flow function $\e(t)$ drop to very low values. During the USR phase, the Hubble flow function $\eta(t)$ also drops from near zero to  \be \lb{etadrop}
\Delta \eta \approx -6.
\ee

Our findings for the CMB scalar tilt $n_s$ with the  fine-tuned parameters $(\a,\x,\f_i)$ are displayed in Fig.~\ref{fig5}
where we also show the dependence of $n_s$ upon $\f_{\rm in}$ and $\theta$.

\begin{figure}[h]
\begin{minipage}[h]{0.49\linewidth}
\center{\includegraphics[width=1\textwidth]{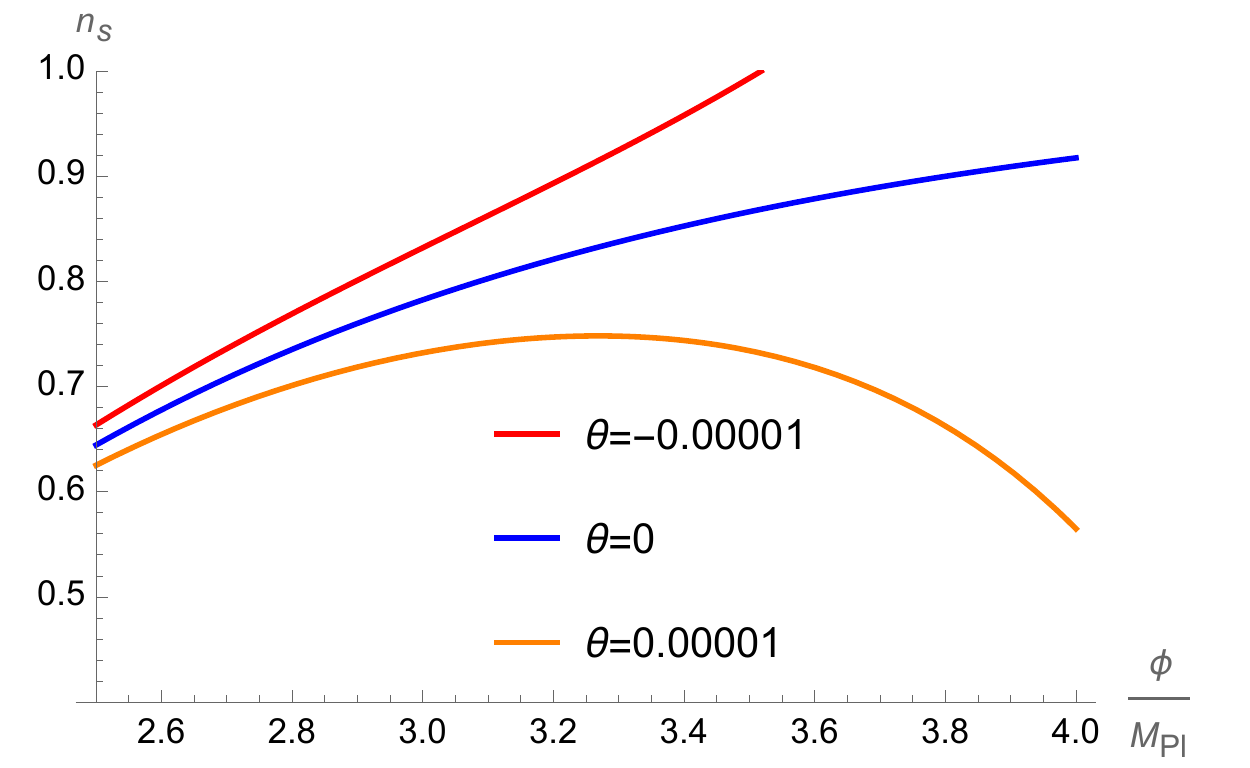}}
\end{minipage}
\hfill
\begin{minipage}[h]{0.49\linewidth}
\center{\includegraphics[width=1\textwidth]{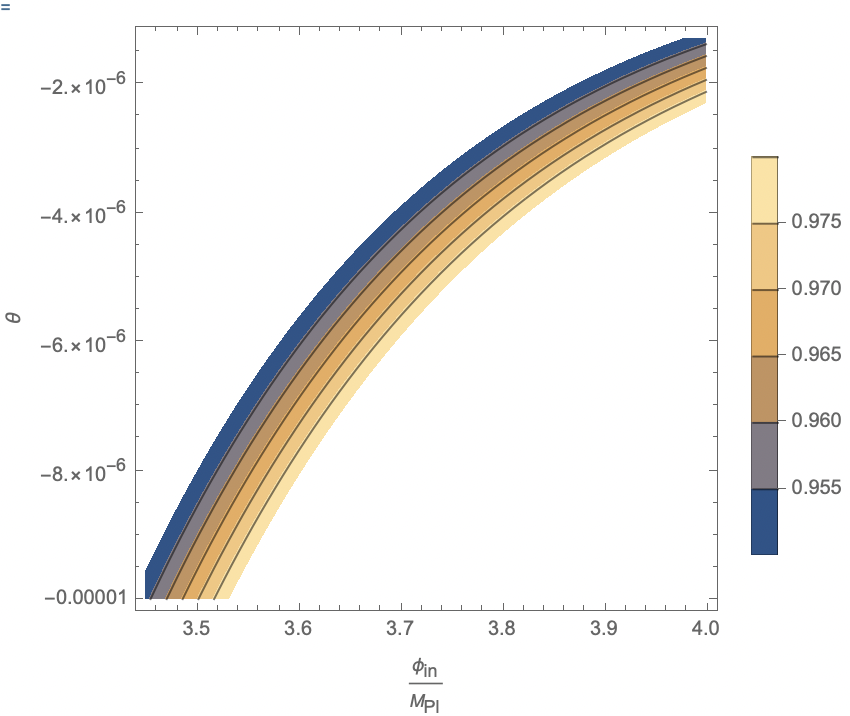}}
\end{minipage}
\caption{The dependence of the CMB scalar tilt $n_s$ upon the inflaton field $\phi$ (on the left), and upon $\theta$ and $\f_{\rm in}$ at the horizon exit  (on the right). The other parameters are $\alpha=0.74265$, $\xi =0.01213$ and $\phi_i=-0.61115$.  }
\label{fig5}
\end{figure}

With the exit value of $\f_{\rm in}/M_{\rm Pl}=3.886$ and the parameter $\theta=-0.000002597$, we find
\be \lb{nsr}
n_s= 0.96498 \quad {\rm and} \quad r=0.03196~,
\ee
which are fully consistent with the CMB measurements (\ref{ns}). This value of $r$  saturates the current CMB upper bound on the
tensor-to-scalar ratio and maximizes the PBH masses (see the next Section).

\section{Power spectrum and PBH formation}

A standard procedure of (numerically) computing the power spectrum $P_R(k)$ of scalar (curvature) perturbations depending upon scale $k$ is based on the Mukhanov-Sasaki (MS) equation \cite{Mukhanov:1985rz,Sasaki:1986hm}. There is a simple analytic formula for $P_R$ in the SR approximation, see e.g., Ref.~\cite{Garcia-Bellido:2017mdw}, which reads
 \begin{equation} \lb{srsp}
 	P_R=\fracmm{H^2}{8 M_{\mathrm{Pl}}^2 \pi^2 \epsilon}~~.
 \end{equation}
We used both approaches in application to our models and found that the difference between the exact results from numerically solving the MS equation and those derived from the SR formula (\ref{srsp}) is small (see Fig.~\ref{fig6} as an example), being irrelevant for our purposes.

\begin{figure}[h]
\begin{minipage}[h]{0.49\linewidth}
\center{\includegraphics[width=1\textwidth]{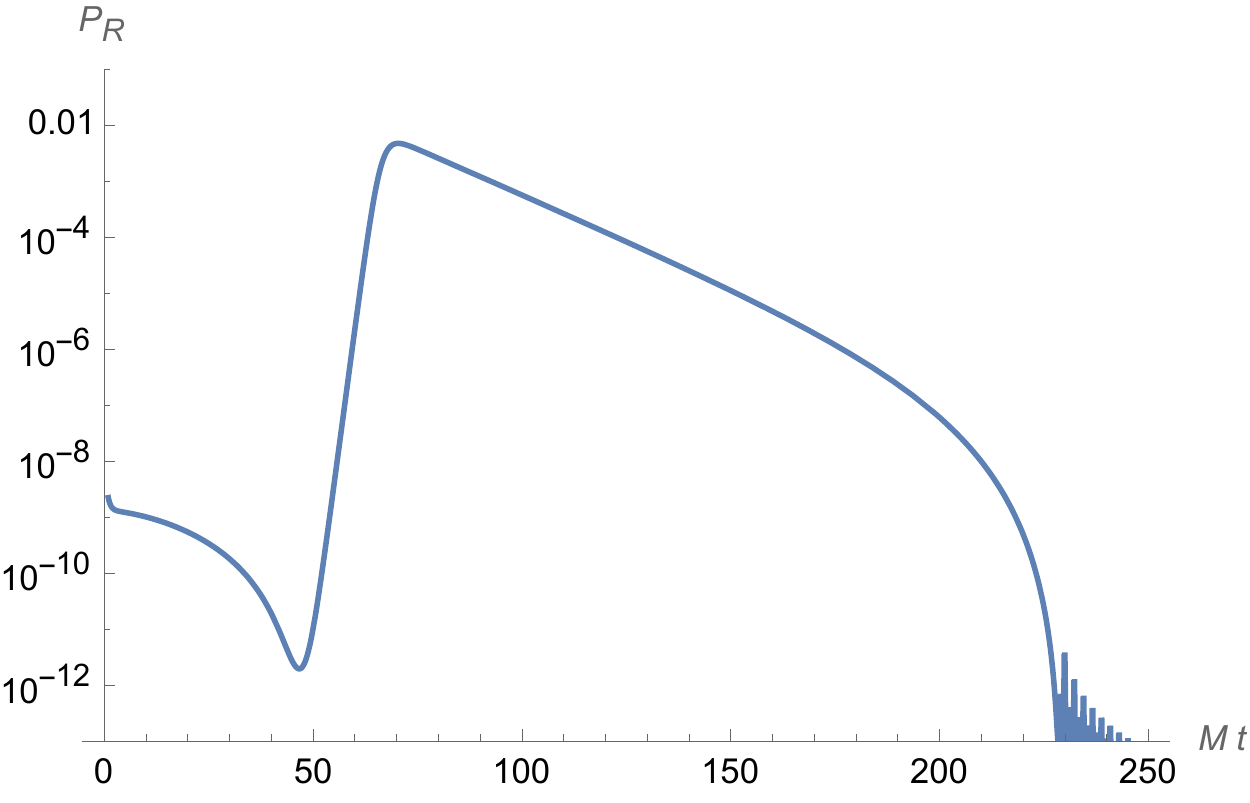}}
\end{minipage}
\hfill
\begin{minipage}[h]{0.49\linewidth}
\center{\includegraphics[width=1\textwidth]{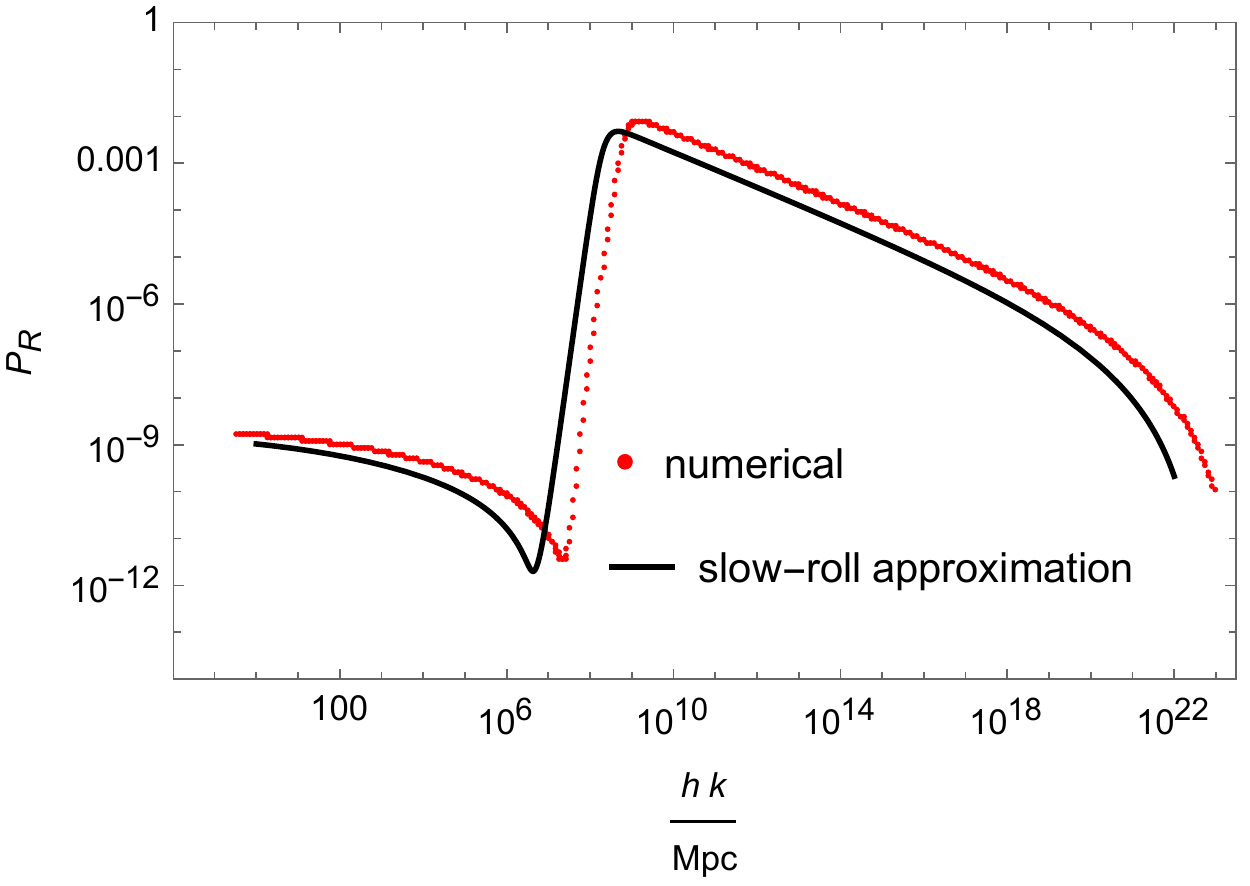}}
\end{minipage}
\caption{The power spectrum of scalar perturbations with the parameters $\alpha=0.74265$, $\xi =0.01213$, $\phi_i=-0.61115$, 
$\theta=-0.000002597$ and  $\phi_{\rm in}/M_{\rm Pl}=3.886$. The peak location is at $k_{\rm peak}=10^9$ Mpc${}^{-1}$.  }
\label{fig6}
\end{figure}

Knowing the location $(k)$ of the peak in the power spectrum allows us to estimate the masses $M(k) $ of the produced PBH as \cite{Carr:2020gox,Escriva:2022duf}
\begin{equation} \lb{pbhm}
	\fracmm{M(k)}{M_{\odot}} \simeq 10^{-16}\left(\fracmm{k}{10^{14} \mathrm{Mpc}^{-1}}\right)^{-2}~~,
\end{equation}
where $M_{\odot}$ is the mass of the Sun,  $M_{\odot}=1.998\cdot10^{33}g$. For instance, with the peak value $k_{peak}\approx 10^9$ Mpc${}^{-1}$  we get 
\be \lb{maxpbhm}
M_{\rm PBH}\approx 10^{-7}M_{\odot}=1.74\cdot10^{26}~g~~,
\ee
which is just between the mass of the Moon, $M_{\rm Moon}= 7.34\cdot 10^{25}$ g and the mass of the Earth, $M_{\bigoplus}=5.97\cdot 10^{27}g$. The PBH mass (\ref{maxpbhm}) should be understood as the {\it maximal\/} PBH mass possible in our classical E-models after extreme fine-tuning of the parameters, keeping agreement with CMB measurements.

As regards other possible values of the CMB tilts $n_s$ and $r$, the squared amplitude $\D^2_{\rm peak}$ of curvature perturbations at the horizon 
exit $r_*=2\p/k_*$, and the related PBH masses according to Eq.~(\ref{pbhm}), derived in our models, they are summarized in Table 1.

\begin{table}[h]
\caption{The values of the CMB tilts $n_s$ and $r$, the squared amplitude $\D^2_{\rm peak}$ of curvature perturbations in the peak, and the PBH masses (\ref{pbhm}) in our single-field inflationary E-models with the potential (\ref{npot}) and the tuned values of the parameters.}
\label{table}
\begin{center}
\begin{tabular}{| c | c | c | c | c | c | c | c |   }
\hline
\phantom{@}$\mathbf{n_s}$\phantom{@} & \phantom{@}$\mathbf{r}$\phantom{@} & \phantom{@}\mbox{\boldmath{$\alpha$}} \phantom{@} & \phantom{@}\mbox{\boldmath{$\theta$}} \phantom{@} &  \phantom{@}\mbox{\boldmath{$\phi_{\textbf{\text{in}}}$}}\phantom{@} &\phantom{@}  $\Delta^2_{\text{peak}}$  \phantom{@} & \phantom{@} $\mathbf{M_{\textbf{\text{PBH}}}}$ \phantom{@}\\ \hline
0.96498 & 0.03196 & 0.74256 & $-$ 0.000002597& 3.886 & 0.008   & $1.7\cdot10^{26}$ g
\\ \hline
0.96494 & 0.03098 & 0.74260 & $-$ 0.000002472 &  3.9 &0.007 & $8\cdot 10^{25}$ g
\\ \hline
0.96496 & 0.01569 & 0.74250 & $-$ 0.000000820 &  4.2 & 0.003 & $5 \cdot 10^{19}$ g \\ \hline

\end{tabular}
\end{center}
\end{table}

The plots of the Hubble flow functions (\ref{Hflow}) in the particular model with the parameters 
$\alpha=0.74265$, $\xi =0.01213$, $\phi_i=-0.61115$, and $\theta=-0.000002597$,  are given in Fig.~\ref{fig7}. Figure \ref{fig7} 
shows the existence of an USR phase with $\abs{\eta}> 1$ between two SR phases, where the USR begins at $Mt_s\approx 46$ and ends at  $Mt_e\approx 70$ with the USR duration of  $M(t_e-t_s)\approx 24$. Our plots in Fig.~\ref{fig7} are similar to those in the literature for the inflationary  models with an USR phase, see e.g., Fig.~1 of Ref.~\cite{Liu:2020oqe} for comparison.

To the end of this Section, we briefly comment on expected one-loop quantum corrections in our E-models, by using the results and
discussions in Refs.~\cite{Kristiano:2022maq,Riotto:2023hoz,Kristiano:2023scm,Riotto:2023gpm,Choudhury:2023rks,Firouzjahi:2023ahg,Franciolini:2023lgy}. 

The Hubble flow function $\eta(t)$ can be approximated by a step function during the USR phase, which led to the (one-loop) perturbative bound on validity of the classical results in the form \cite{Kristiano:2022maq,Kristiano:2023scm} 
\begin{equation}\label{oneloop}
	\frac{1}{4}(\Delta\eta)^2\left(1.1+\log{\fracm{k_e}{k_s}}\right)\Delta^2_{{\rm peak}}\ll 1~~,
\end{equation}

\begin{figure}[h]
\begin{minipage}[h]{0.44\linewidth}
\center{\includegraphics[width=1\textwidth]{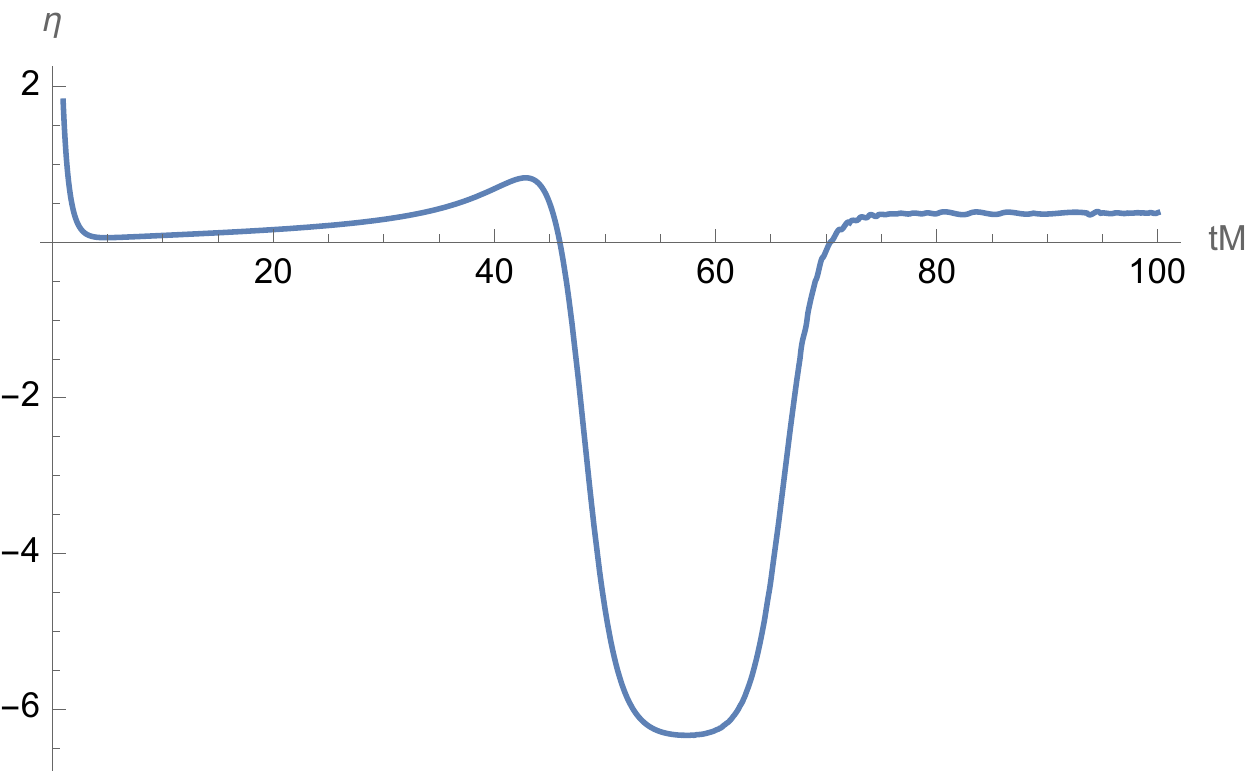}}
\end{minipage}
\hfill
\begin{minipage}[h]{0.44\linewidth}
\center{\includegraphics[width=1\textwidth]{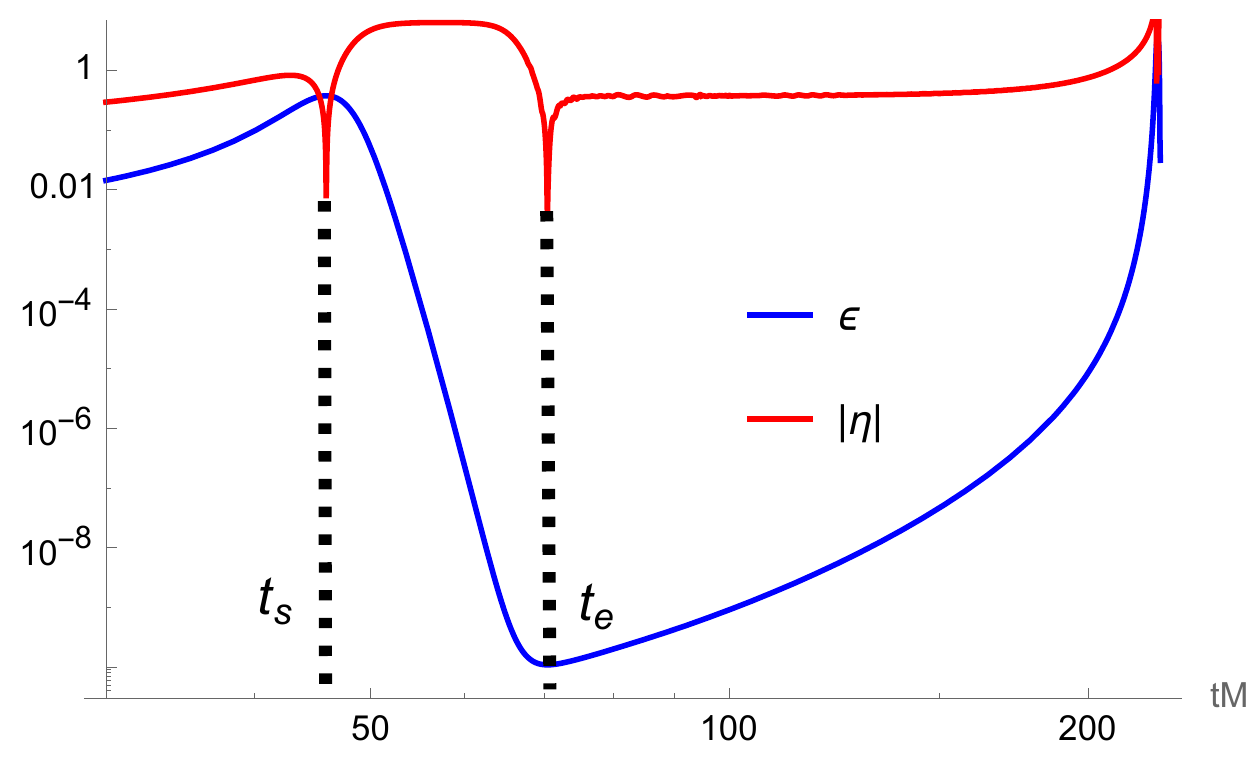}}
\end{minipage}
\caption{The Hubble flow functions (\ref{Hflow}) for the parameters $\alpha=0.74256$, $\xi =0.01213$, $\phi_i=-0.61115$ and $\theta=-0.000002597$.  The other related values are $\phi_{\rm in}/M_{\rm Pl}=3.886$, $k_{s}\approx 4.4\cdot10^6$ Mpc${}^{-1}$, $k_{e}\approx 3\cdot 10^8$ Mpc${}^{-1}$ and $\Delta\eta=-6.3$. Both plots show the USR phase for $\abs{\eta}>1$. The plot (right) is the superposition of the Hubble flows for $\e(t)$ and $\abs{\eta}(t)$.}
\label{fig7}
\end{figure}

\noindent where the amplitude squared of scalar curvature perturbations was taken in Refs.~\cite{Kristiano:2022maq,Kristiano:2023scm} of the order 
$\Delta_{\rm peak}^2\sim {\cal O}(10^{-2})$. Given $(\Delta\eta)^2\approx 36$, Eq.~(\ref{oneloop}) implies the perturbative bound on the power spectrum as \cite{Kristiano:2022maq,Kristiano:2023scm}
\be \lb{bound} 
 \Delta^2_{{\rm peak}}\ll 0.055 \sim {\cal O}(10^{-2})
\ee
that might rule out classical results about PBH production in any single-field model of inflation with  a near-inflection point. In our models we have the smooth function $\eta (t)$ and $\Delta\eta\approx -6.3$ that implies a stronger bound,
\be \lb{bound1} 
 \Delta^2_{{\rm peak}}\ll 0.025 \sim {\cal O}(10^{-2})~~.
\ee
However, the values of  $\Delta^2_{\rm peak}$ given in Table 1 are smaller (though not much smaller) than the bound (\ref{bound1}).  There is an uncertainty in the value of $\Delta_{\rm peak}^2\sim {\cal O}(10^{-2})$ used in Refs.~\cite{Kristiano:2022maq,Kristiano:2023scm} for PBH production, which is related to an uncertainty in the smoothed density contrast constant and relaxes the peak value (needed for PBH formation) to  $\Delta_{\rm peak}^2\sim {\cal O}(10^{-3})$ \cite{Riotto:2023gpm}. Therefore, our classical results obtained in the fine-tuned E-models for PBH production are still not ruled out by quantum corrections, at least as regards the bottom line in Table 1.

\section{Conclusion}

A production of PBH with the masses beyond the Hawking (black hole) radiation limit of $10^{15}$ g in the early Universe is not a generic feature of single-field models of inflation with a near-inflection point because it requires fine-tuning of the parameters defining the inflaton potential and depends upon duration of the SR and USR phases. We adopted the maximal fine-tuning with the minimal number of parameters in the (generalized) E-type $\a$-attractor models of inflation, while keeping agreement with CMB measurements. 

We numerically scanned the parameter space of our models in order to find the parameter values leading to a significant enhancement (by $10^6$ to $10^7$ times) of the power spectrum of scalar (curvature) perturbations. We confirmed high sensitivity of our results to a choice of the parameters, in agreement with general expectations \cite{Cole:2023wyx}. 

Compared to Ref.~\cite{Frolovsky:2022qpg}, we added the new parameter $\theta$ at the new term (proportional to $y^{-2}$) in the inflaton scalar potential in order to reach perfect agreement with the observed CMB values of the cosmological tilts $n_s$ and $r$.

Fine-tuning of the parameters in our models was needed not only to match the CMB measurements (inflation is robust in the E-models) but also to generate PBH production at lower scales during the USR phase, which would lead to a sizable portion of PBH in the current dark matter. The equations for the PBH masses in the literature \cite{Carr:2020gox,Escriva:2022duf,Karam:2022nym} are sensitive to a detailed
shape of the peak in the power spectrum and the USR phase duration, so that in this paper we only used the rough estimate in Eq.~(\ref{pbhm}). Taking into account those uncertainties may change the PBH masses by 1-2 orders of magnitude. 

Though a PBH production with the masses of approximately $10^{26}$ g was found to be possible in the classical theory (see the first row in Table 1), those PBH  may be ruled out by quantum corrections. As regards the asteroid-size PBH with the masses of approximately $10^{19}$ g, which are possible candidates for the whole dark matter in the Universe \cite{Carr:2020gox,Escriva:2022duf,Frolovsky:2022qpg}, the quantum (one-loop) corrections were shown to be suppressed by one order of magnitude according to the estimate in Eq.~(\ref{bound1}) and the  third row of Table 1. Hence, our generalized E-models of inflation and PBH production are still not ruled out.

\section*{Acknowledgements}

This work was partially supported by Tomsk State University under the development program Priority-2030.  SVK was also supported by Tokyo Metropolitan University, the Japanese Society for Promotion of Science under the grant No.~22K03624, and the World Premier International Research Center Initiative, MEXT, Japan.

\bibliography{Bibliography}{}
\bibliographystyle{utphys}

\end{document}